\documentstyle[aps,prb,multicol,epsf]{revtex}

\begin{document}
\draft

\title{
Periodic orbit resonances 
in layered metals in  tilted magnetic fields\\
}

\author{Ross H. McKenzie\cite{email}
 and Perez Moses}

\address{School of Physics, University of New
South Wales, Sydney 2052, Australia}

\date{Received April 1, 1999}

\maketitle

\begin{abstract}
The frequency dependence of the interlayer
conductivity of a layered Fermi liquid in a magnetic
field which is tilted away from the normal to the layers is considered.
For both quasi-one- and quasi-two-dimensional
systems
resonances occur when the frequency is a harmonic of
the frequency at which the magnetic field causes
the electrons to oscillate on
the Fermi surface within the layers.
The intensity of the different harmonic resonances
varies significantly with the direction of the field.
The resonances occur for both coherent and weakly incoherent
interlayer transport and so their observation does 
not imply the existence of a three-dimensional Fermi surface.
\\
\\
To appear in {\it Physical Review B, Rapid Communications}, October 15,
1999.
\\
\end{abstract}



\begin{multicols}{2}

There is considerable interest in performing
frequency-dependent transport measurements 
on strongly correlated metals in the hope that
they will provide 
new information about the metallic state
such as a direct determination of the scattering rate.
Layered organic metals\cite{ish,wos} are ideal for such
experiments due to their high purity
and a number of experiments have been performed.\cite{expt,ardavan}
In this paper we show that 
when the magnetic field is tilted away from the normal
to the layers that there are well-defined resonances
in the interlayer conductivity when the frequency
equals a harmonic of the frequency at which the
magnetic field causes electrons to traverse the Fermi surface
within the layers. This occurs for both quasi-two and
quasi-one-dimensional systems. The intensity of the
resonances at different harmonics varies significantly
with the direction of the field.
For example, it is possible to choose the field 
direction so one will see predominantly only odd or 
even harmonic resonances.
In general, a three-dimensional Fermi surface is not
necessary for the observation of the resonances.
We also compare our results to
previous theoretical work\cite{hill1,blundell} 
which has involved more
complicated band structures.


We assume a Fermi liquid within the layers
with the simplest possible dispersion relation
  $\epsilon(k_x,k_y)$.
For quasi-one-dimensional systems we take
\begin{equation}
  \epsilon(k_x,k_y)=
\hbar v_F (|k_x| - k_F) - 2 t_b \cos (k_y b) 
\label{disp0}
\end{equation}
where $v_F$ is the Fermi velocity, $k_F$ is the
Fermi wave vector, $t_b$ the interchain hopping integral,
and $b$ the interchain distance.
For the quasi-two-dimensional case,
   $\displaystyle \epsilon(k_x,k_y) ={\hbar^2 \over 2 m^*} (k_x^2 + k_y^2) $ 
 where $m^*$ is the effective mass.
Solution of the semi-classical equations of motion shows
that in a magnetic field $B$ normal to the layers electrons
move across the Fermi surface within the layers
at a periodic orbit frequency $\omega_0$, which equals
  $ \displaystyle {e v_F b B   \over \hbar } $  
or
$ \displaystyle {e  B   \over m^*   } $ 
for the quasi-one- and quasi-two-dimensional cases, respectively.

We consider a magnetic field tilted at an angle $\theta$
away from the normal to the layers.
For the 
 quasi-one-dimensional case we at first only
consider the case where the field is confined
to the $x-z$ plane, i.e., the plane containing the 
most and least-conducting directions.
This is done for reasons of simplicity;
later in the paper, we consider more general field 
directions for the quasi-one-dimensional case.
We have calculated the frequency-dependent interlayer conductivity
for two different models for
the interlayer transport: coherent and 
weakly incoherent interlayer transport. 
A detailed discussion of these two models
is given in Reference \onlinecite{moses2}.
The former model is associated with a three-dimensional Fermi
surface. The latter 
occurs when the intralayer scattering rate $1/\tau$
is much larger than the interlayer
hopping integral $t_c$ and so one cannot define 
a wave vector perpendicular to the layers.
In that case the Fermi surface is only defined within 
the layers.
In this paper we present the results of  our
calculations and discuss their implications.
The details of the calculations are similar to those
for the dc conductivity.\cite{moses2}
The scattering time $\tau$ is assumed to be the same 
at all points on the Fermi surface.
We find that
the frequency-dependent interlayer conductivity for {\it both}
coherent (except when the field is very close
to the layers) and weakly incoherent interlayer transport is,
for a frequency $\omega$,
\begin{equation}
{\sigma_{z z}(\omega) \over \sigma_{z z}^{0}} = 
\sum_{\nu=-\infty}^{\infty} { J_{\nu}(\gamma \tan\theta)^2
\over {1 + (\omega - \nu \omega_0 \cos \theta)^2 \tau^2 } }
\label{amro}
\end{equation}
where $J_{\nu}(x)$ is the $\nu$-th order
Bessel function, $\omega_0$ is the oscillation frequency
associated with the magnetic field, and 
$\sigma_{z z}^{0}$ is the dc conductivity in the absence
of a magnetic field. 
$\gamma $ is a constant that depends on the geometry of
the Fermi surface; it is
  $ \displaystyle{2 t_b c  \over \hbar v_F } $
  and $ k_F c$  for the quasi-one- and quasi-two-dimensional
cases, respectively.

{\it AMRO.}
In the dc limit ($\omega=0$),
Eqn. (\ref{amro}) reduces to our result for the dc
conductivity.\cite{moses2,moses}
At a fixed field, the interlayer resistivity
oscillates as a function of the tilt angle;
referred to as angular-dependent magnetoresistance
oscillations (AMRO).\cite{wos}
These oscillations are known as the Yamaji and Danner
oscillations for the quasi-two- and quasi-one-dimensional 
cases, respectively.
The interlayer dc resistivity is a local maximum 
at the   angles,  
$\theta = \theta^m_{max}$ where
\begin{equation}
\gamma \tan\theta^m_{max} = \left(m - { 1\over 4} \right) \pi \  \ 
\ \ m=1,2,3, \cdots.
\end{equation}
and a local minimum at angles, $\theta = \theta^m_{min}$ where
\begin{equation}
\gamma \tan\theta^m_{min} = \left(m + { 1\over 4} \right) \pi \  \ 
\ \ m=1,2,3, \cdots.
\end{equation}

{\it Drude peak.}
In zero field ($\omega_0=0$)  or when the
field is perpendicular to the layers ($\theta=0$)
(\ref{amro}) reduces to the Drude form,
\begin{equation}
\sigma_{z z}(\omega) =          
{\sigma_{z z}^{0}  
 \over {1 + (\omega \tau)^2 } }
\label{drude}
\end{equation}
It is interesting that this result holds for 
weakly incoherent transport.
This means that
 even when the intralayer scattering
rate is so large ($1/\tau \gg t_c)$ that the interlayer mean-free path
is much less than the layer spacing one
will still observe a Drude peak when there is 
a Fermi liquid within each of the layers.

{\it Periodic orbit resonances.}
The most important property of (\ref{amro}), and the focus
of this paper, is that the frequency-dependent conductivity
has resonances  at harmonics of the periodic orbit frequency,
\begin{equation}
\omega = n \ \omega_0 \cos \theta
\label{res}
\end{equation}
where $n$ is an integer. 
The $n-$th resonance has  intensity
$J_{n}(\gamma \tan\theta)^2$, which varies significantly with
the orientation of the magnetic field.

What is the physics behind these resonances?
In the coherent case the interlayer electronic group
velocity for a trajectory on the Fermi surface starting
at $k_z(0)$ is
\begin{eqnarray}
v_z(k_z(0), t) \sim  
 \sin(\gamma \tan \theta \sin (\omega_0 \cos \theta t) + k_z(0) c)
\nonumber \\
\sim  \sum_n J_n(\gamma \tan \theta) \sin (n \omega_0 \cos \theta t)
\label{vel}
\end{eqnarray}
which will produce an alternating current
at harmonics of the periodic orbit frequency.
An ac electric field in the $z$ direction will have
resonances with this current.
For the weakly incoherent case the overlap of the
time-dependent
wavefunctions between neighbouring layers has the
same form as the right hand side of (\ref{vel}).

In a typical experiment the layered metal is
placed inside a microwave cavity which has a
fixed resonance frequency $\omega$.
The magnetic field is then varied and one looks
for resonances in the cavity response,
which will generally be dominated by changes in
the interlayer resistivity. Hence,
in Fig. 1 we plot
$\sigma_{z z}(\omega) $ versus $\omega_0$ which
is proportional to the magnetic field.
The figure shows a very rich structure in the 
field dependence in a tilted field,
with resonances at many different harmonics
of the periodic orbit frequency $\omega_0 \cos \theta$.
Furthermore, the intensity of the different
resonances is very different, depending on whether
the field is at an angle corresponding to
an AMRO maximum or minimum.
For example, at the first AMRO minimum the first
harmonic is suppressed whereas at the fifth
AMRO maximum the second and fourth harmonics are
suppressed.
For $z > n$, the asymptotic form $J_{n}(z) \sim  \sqrt{2\over \pi z}
\cos \left( z - { n \pi \over 2} -
{\pi \over 4}\right)$ holds.
Hence, in general,
if the field is tilted at the $m$'th AMRO maxima (minima) one will see
only the odd (even) harmonics with $n < m.$
To illustrate the above discussion
in Fig. 2 we plot the intensity of the resonance 
at the $n$-th harmonic as a function of $n$
at several different angles.
For example, this explains the absence of
the $n=$ 2, 4, and 7 peaks for
the fifth AMRO maximum.  The
peak at the lowest field in Fig. 1 represents a
superposition of resonances with the
$n=$ 12, 13, and 14 harmonics; they cannot be resolved
because $\omega_0 \tau \sim 1$.

The richness of the response shown in Fig. 1
has important implications for the interpretation
of experimental results, many of which contain
multiple resonances.
As emphasized previously by Hill\cite{hill1}, one should be cautious 
about assigning different resonances to
different bands with different effective masses.
It is quite possible that if the sample is aligned
(perhaps inadvertantly) so that the field is tilted
away from the normal to the layers
that the different features observed actually correspond
to harmonics of a single band.

{\it Open orbit resonances.}
In a quasi-two-dimensional metal the
presence of periodic orbits is clearly
seen in magnetic oscillations such as the
de Hass - van Alphen effect.
In contrast, the periodic orbits in a
quasi-one-dimensional metal do not
produce such oscillations, raising the question
of what might be clear experimental
signatures of the existence of the periodic orbits.\cite{caveat}
The result (\ref{amro}) shows that
frequency-dependent
resonances in the interlayer conductivity
provide a means to directly establish the
periodic motion of the electrons in a quasi-one-dimensional
system and to determine the Fermi velocity $v_F$.
It has been proposed that in strongly correlated
chains with weak interchain coupling that
coherence is confined to the chains.\cite{str}
One would not expect periodic orbit behavior
within the layers in such circumstances.
Hence, detection of periodic orbit resonances in a particular material
is a possible way to rule out this proposal\cite{str}
for that material.

Gorkov and Lebed\cite{gor} have previously proposed
a method of detecting periodic orbit resonances
in a quasi-one-dimensional metal using
singularities in the surface impedance.
However, Hill\cite{hill1} pointed out that typical experiments
do not satisfy the requirement that the skin depth be
less than the cyclotron radius and mean-free path.
The method here involves the bulk conductivity
and so is not under similar constraints.
Hill\cite{hill1} considered the 
frequency-dependent interlayer conductivity for
a three-dimensional band structure where the interlayer
hopping depends on the wave vector within the layer
and the intralayer Fermi surface is elliptical.
He found that resonances associated with the
second and third harmonics of the cyclotron frequency
were possible. The present work shows that some
of the effects he found associated with a complicated
band structure can also be produced with a simple band
structure in a tilted field.

{\it Kohn's theorem.}
Some of the motivation for frequency-dependent measurements
in layered metals has been the idea that
cyclotron resonance measurements can provide
information about the strength of many-body
effects which lead to differences between
the effective mass deduced from cyclotron resonance and 
that deduced from magnetic  oscillations.
Kohn\cite{kohn} showed
that for an {\it isotropic}
three dimensional metal with the electric field {\it perpendicular}
to the magnetic field, a cyclotron resonance experiment
will measure a resonance frequency which is independent of
the strength of the electron-electron interactions.
In contrast, the effective mass determined from
the temperature dependence of magnetic oscillations {\it is}
renormalised by the interactions.
Hence, a comparison of the effective mass from 
the two measurements provides a means
to determine the importance of electron-electron interactions
in a material. 
Kohn's argument easily follows through for a
two-dimensional electron gas with the magnetic field
perpendicular to the plane of the 2DEG and the 
oscillating electric field parallel to  the plane.
However, the resonances in the interlayer conductivity discussed above
are determined by the component of the magnetic
field perpendicular
to the layers, which is {\it parallel} to the
direction of the  oscillating electric field.
One finds in that case that Kohn's argument
breaks down and so without further work it is not
clear whether       Kohn's  ideas are applicable
to the interlayer conductivity of layered metals.
In particular, it is quite possible that the
resonances discussed here include all   the
effects of the interactions.

{\it Typical parameter values.}
Typical microwave cavities operate at 
$\omega=$ 30-150 GHz and to observe the resonances discussed
here we should have at least $\omega \tau >    3$.
This means the sample must be of sufficient purity
that the scattering time $\tau >   2 \times 10^{-11}$ sec.
The values for quasi-two-dimensional 
BEDT-TTF materials deduced from the Dingle temperature ($\sim 1$ K)
for magnetic oscillations are typically
about $10^{-11}$ sec. The Dingle temperature
tends to underestimate the true value and so with the best samples
it should be possible to detect the effects discussed here.
For the quasi-one-dimensional
(TMTSF)$_2$ClO$_4$
Danner, Kang and Chaikin\cite{dan} estimated
$\tau=  4 \times 10^{-12}$ sec from comparing
the angular-dependent interlayer magnetoresistance
with semi-classical calculations.
If $\omega= 100$ GHz, 
this would give $\omega \tau \sim 0.4$ and
so one would need a better sample or to work in the
infra-red to see a clear resonance.
Furthermore, to detect the first harmonic
the field needs to be tilted near the
first AMRO maximum, $\theta_{max}^1= 84 \deg $.
If $\omega= 100$ GHz, the first harmonic will occur at
a field $B = \hbar \omega/(e v_F b \cos \theta_{max})= $ 2 T.

{\it Quasi-one-dimensional 
systems with the field in a general direction.}
For the field 
$\vec{B}=(\sin \theta \cos \alpha, \sin \theta \sin \alpha, \cos \theta)$,
the interlayer conductivity is
\begin{equation}
{\sigma_{z z}(\omega) \over \sigma_{z z}^{0}} = 
\sum_{\nu=-\infty}^{\infty} { J_{\nu}(\gamma \tan\theta \cos \alpha)^2
\over {1 + (\omega - \omega_0
 (\nu \cos \theta - c \sin \theta \sin \alpha /b 
))^2 \tau^2 } }
\label{amro3}
\end{equation}
which reduces to (\ref{amro}) when $\alpha = 0$.\cite{aside}
This can used to  interpret
recent observations\cite{ardavan} of frequency-dependent resonances
 in the low-temperature
phase of $\alpha$-(BEDT-TTF)$_2$KHg(SCN)$_4$.
In order to make a clearer connection
To do this we change to      
angular co-ordinates similar to that used
in Ref.          \onlinecite{ardavan},
where     
$\vec{B}=(\sin \phi, \cos \phi \sin \psi, \cos \phi \cos \psi)$.
The $n$-th resonance will occur with intensity
$J_n(\gamma \tan \phi / \cos \psi)^2$
when
$ \omega /\omega_0 =
 n \cos \phi \cos \psi
-(c/b) \cos \phi \sin \psi $
which can be re-written as
$ {\omega \cos \psi_n \over \omega_0 \cos \phi} = n \sin (\psi + \psi_n)$
where $\tan \psi_n = c /n b$ and $\omega_0 \cos \phi$ is proportional
to $B_\parallel$, the component of the field parallel
to the $y-z$ plane.
This is in a similar form to Eqn. (6) in Ref. \onlinecite{ardavan}.
If  $b/c = 1.3 $ then
$\psi_{1} = $ 38 or 142  degrees
and $\psi_2 = 21$ degrees.
These values are in reasonable agreement with the
values deduced in Ref. \onlinecite{ardavan}
from the two resonances observed in the experiment.

An alternative interpretation of the origin of
the resonances was given in 
Ref. \onlinecite{ardavan} in terms of a
quasi-one-dimensional band structure,\cite{osada}
where the  interlayer transport includes hopping beyond
next-nearest neighbours,
\begin{equation}
  \epsilon(k_x,k_y,k_z)=
\hbar v_F (|k_x| - k_F) - \sum_{m,n} t_{mn} \cos (m k_y b + n k_z c) 
\label{disp1}
\end{equation}
and each sum is over all the integers.
This can be compared to our dispersion (\ref{disp0})
which corresponds to (\ref{disp1}) with the only non-zero elements
being $t_{10} = t_b $ and $t_{01} = t_c $, and
we do not necessarily assume coherent interlayer transport.
Blundell, Ardavan, and Singleton have shown\cite{blundell}
that for a magnetic field in the $y-z$ plane
the frequency-dependent interlayer conductivity
is
\begin{equation}
{\sigma_{z z}(\omega) \over \sigma_{z z}^{0}} = 
\sum_{m,n} { ( n t_{mn}/t_c)^2
\over {1 + (\omega - \omega_0 (n \cos \theta - m c \sin \theta/b))^2 \tau^2 } }
\label{amro2}
\end{equation}
which has resonances when
$\omega  = \omega_0 (n \cos \theta - m c \sin \theta/b)$
where $m$ and $n$ are integers.
The intensity of these resonances depends on
the hopping matrix elements, $ t_{mn}$.
Whereas in our treatment these features arise from
a component of the field in the $x$-direction, in Ref.
\onlinecite{ardavan} they are attributed to the
presence of warping in the Fermi surface associated
with $t_{11}$ and $t_{12}$.\cite{skeptic}
Also, since (\ref{amro3}) holds for both coherent and
weakly incoherent interlayer transport
 even the existence of a three-dimensional Fermi surface
is not necessary for the observation of frequency-dependent
resonances.
Provided there is a Fermi surface within the layers and the
intralayer momentum is conserved in transport
between adjacent layers then the resonances are observable.

This work    was supported by the Australian Research Council.
We thank S. Hill,  M. Naughton, L. Balicas, and J. Singleton
 for very helpful discussions.

\narrowtext

\vskip -0.5cm
\begin{figure}
\centerline{\epsfxsize=8.3cm \epsfbox{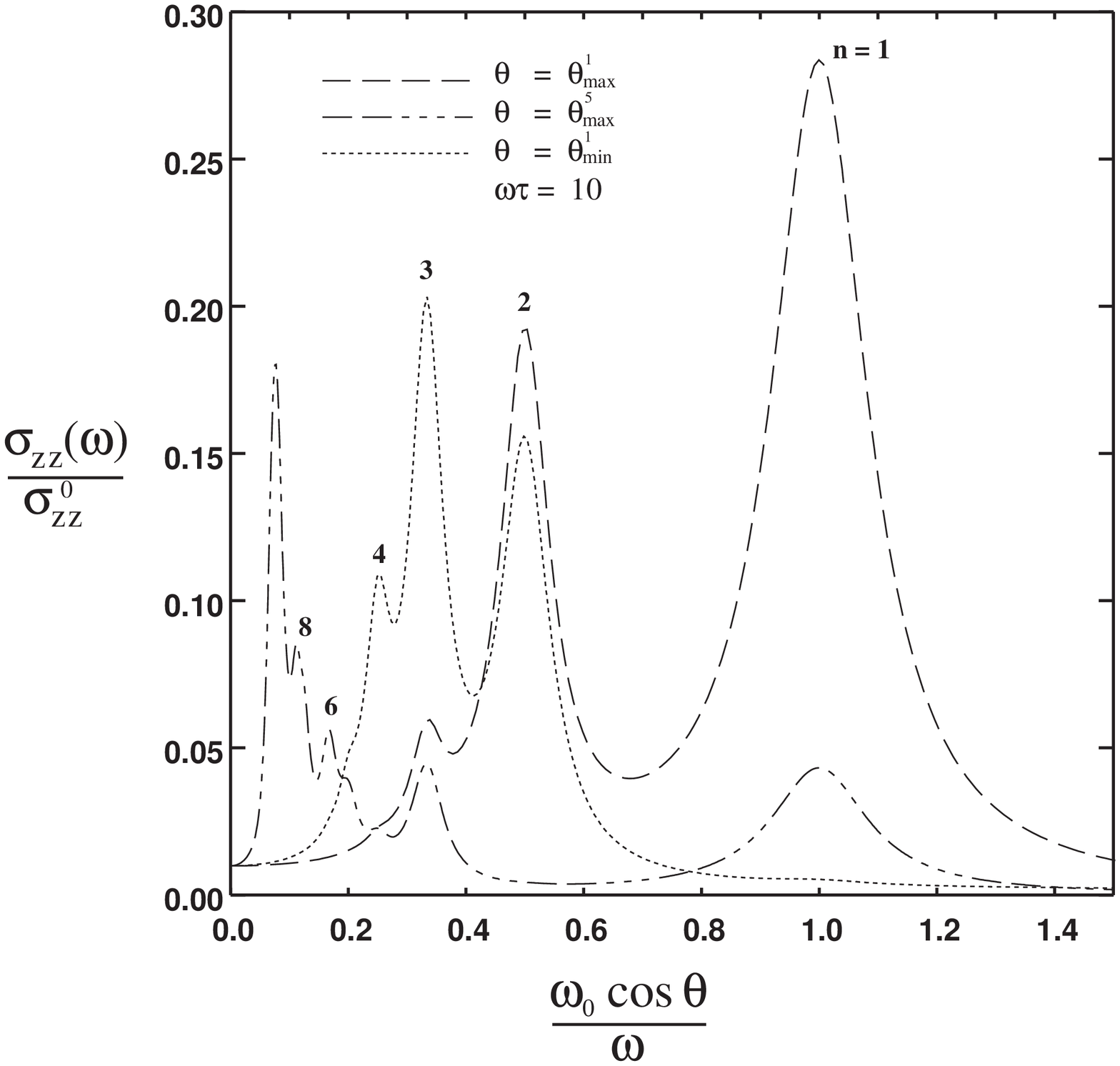}}
\caption{
Magnetic field dependence of the interlayer conductivity
at a fixed frequency $\omega$ when the 
magnetic field is tilted at
several different angles $\theta$ away from the normal to the layers.
The integers given near the peaks denote at which
harmonic of the periodic orbit frequency $\omega_0 \cos \theta$
at which the resonance occurs.
Note that if the field is tilted at
an angle corresponding to the $m$'th AMRO maxima (minima) peaks only  
occur at the odd (even) harmonics with $n < m$.
The ac conductivity $\sigma_{zz}(\omega)$
is normalised to the dc conductivity
in zero field, $\sigma_{zz}^0$.
The scattering time $\tau$ is such that $\omega \tau = 10$.
\label{fig1}}
\end{figure}

\begin{figure}
\centerline{\epsfxsize=8.3cm \epsfbox{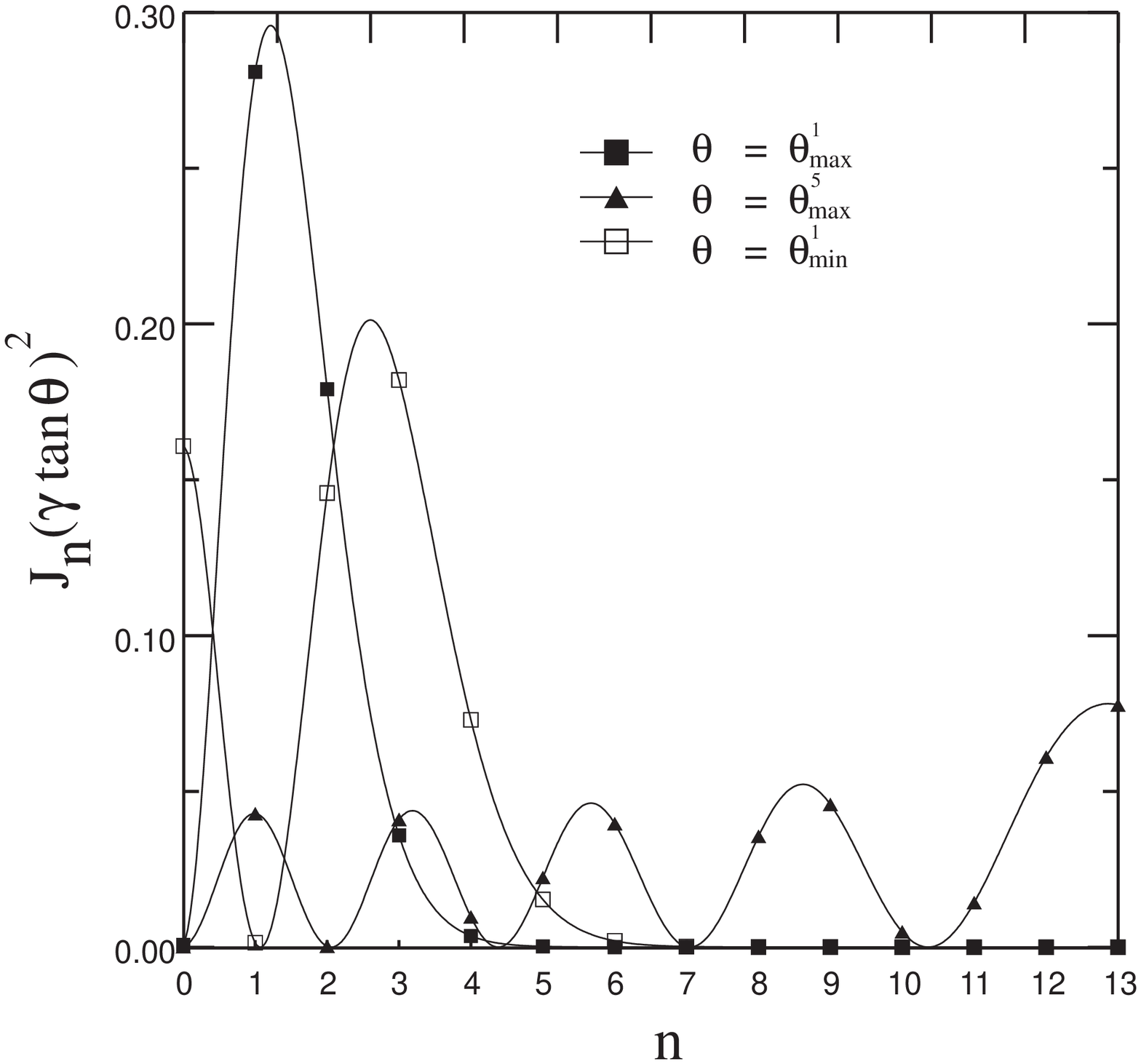}}
\caption{
The intensity of the different harmonic resonances in the interlayer
conductivity the field is tilted
at angle corresponding to different AMRO maxima and minima.
\label{fig2}}
\end{figure}

\end{multicols}

\begin{references}
\bibitem{skeptic}
Generally on the grounds of quantum chemisty
one would expect $t_{12} \ll t_{11} \ll t_{10} = t_c$
and so it is appealing that it is unnecessary to assign them comparable
values in order to explain experimental results.
\end{references}
\end{document}